\documentclass[a4paper]{jpconf}
\usepackage{amsmath}           %allows the use of align
\usepackage{graphicx}
\usepackage[separate-uncertainty,multi-part-units = single, allow-number-unit-breaks]{siunitx}
\usepackage{subfig}

\newcommand{\figref}[1]{Figure~\ref{#1}}    % Standard
\newcommand{\figrefbra}[1]{Fig.~\ref{#1}}   % In brackets 
\newcommand{\Figref}[1]{Figure~\ref{#1}}    % Beginning of a sentence
\newcommand{\secref}[1]{Section~\ref{#1}}   % Standard
\newcommand{\Secref}[1]{Section~\ref{#1}}   % Beginning of a sentence
\begin{document}
\title{Commissioning of a hybrid readout TPC test set-up and gas gain simulations}

\author{A.~Deisting}

\address{Royal Holloway, University of London, Egham Hill, Egham, TW20 0EX, UK}

\ead{alexander.deisting@cern.ch}

\begin{abstract}
A hybrid readout Time Projection Chamber (TPC) has a simultaneous optical- and charge readout. The optical readout provides 2D images of particle tracks in the active volume, whilst the charge readout provides additional information on the particle position perpendicular to the image plane. A hybrid readout TPC working at high pressure is an attractive device for physics cases where an excellent space point resolution and a high target density is required as \textit{e.g.} measuring a neutrino beam at the source of a long baseline neutrino oscillation experiment. In this paper we present two different lines of work towards the goal of developing hybrid TPC technology: a) Commissioning of a set-up with gas electron multipliers employing optical and charge readout. b) An analytical parametrisation of the gas gain for a multi wire proportional chamber based on \textsc{garfield++} simulations, which -- when validated with measurements -- allows to skip these simulations in the future altogether.
\end{abstract}

\section{Introduction}

Time Projection Chambers (TPCs) are heavily relied on as tracking detectors, be it at collider experiments ({\it e.g.} ALICE: \cite{alme2010alice}), in experiments searching for Dark Mater (DM) ({\it e.g.} DarkSide 20k: \cite{Aalseth:2017fik}) or in neutrino oscillation experiments (\textit{e.g.} the Deep Underground Neutrino Experiment (DUNE): \cite{Abi:2020wmh} and other volumes). The gas or liquid volume of a TPC serves as a continuous detection medium and as a scattering target for \textit{e.g.} DM or neutrinos.\\
Gas filled TPCs have a low momentum threshold for particle detection: A proton, {\it e.g.}, with a kinetic energy of \SI{5}{\mega\electronvolt} (\SI{40}{\mega\electronvolt}) will leave a \SI{1}{\centi\meter} track in a TPC filled with gaseous (liquid) argon. Highly granular readout is needed to sample a few points along its track. A specific example for the need of resolving a $\sim\!\!\SI{1}{\centi\meter}$ proton track is the measurement of the neutrino scattering cross section on nuclei in a gas or a liquid. These measurements are needed to address one of the leading sources of systematic uncertainties in neutrino oscillation experiments: Modelling of Final State Interactions (FSIs) \cite{Alvarez-Ruso:2017oui}. FSIs occur when particles are ejected from a nucleus after the neutrino scattering. A gas filled TPC offers the opportunity to measure final state particles (FSPs) at low momenta and is thus capable of resolving the differences between FSI models used by \textit{e.g.} GENIE \cite{Andreopoulos:2015wxa} and NEUT \cite{Hayato:2002sd}. Ironing out these discrepancies allows to tackle the dominant systematic uncertainty faced when analysing the results of neutrino oscillation measurements.

To enhance the event rate for neutrino-nucleus scattering in gas filled TPCs, the pressure of the gas and thus the target density can be increased. For the DUNE near detector a high pressure gas TPC is foreseen to characterise the neutrino beam before it undergoes oscillation and to measure neutrino-argon scattering cross-sections, which entails measuring FSPs \cite{abud2021deep}.

In this paper we will introduce the concept of hybrid optical and charge readout of a TPC in \secref{sec:concept:hybrid}. First steps to constructing an optical TPC prototype are discussed, too. \Secref{sec:simulations} discusses efforts to simulate the charge amplification factor in an ALICE ReadOut Chamber (ROC) and to parametrise it as a function of applied voltages and gas mixture in order to avoid simulating these quantities in the future.

\section{Time projection chambers with hybrid optical and charge readout}
\label{sec:concept:hybrid}

\begin{figure}
\centering
\subfloat[]{\label{sec:concept:fig:otpc:concept}\includegraphics[width=0.42\columnwidth, trim= 0 100 0 450, clip=true]{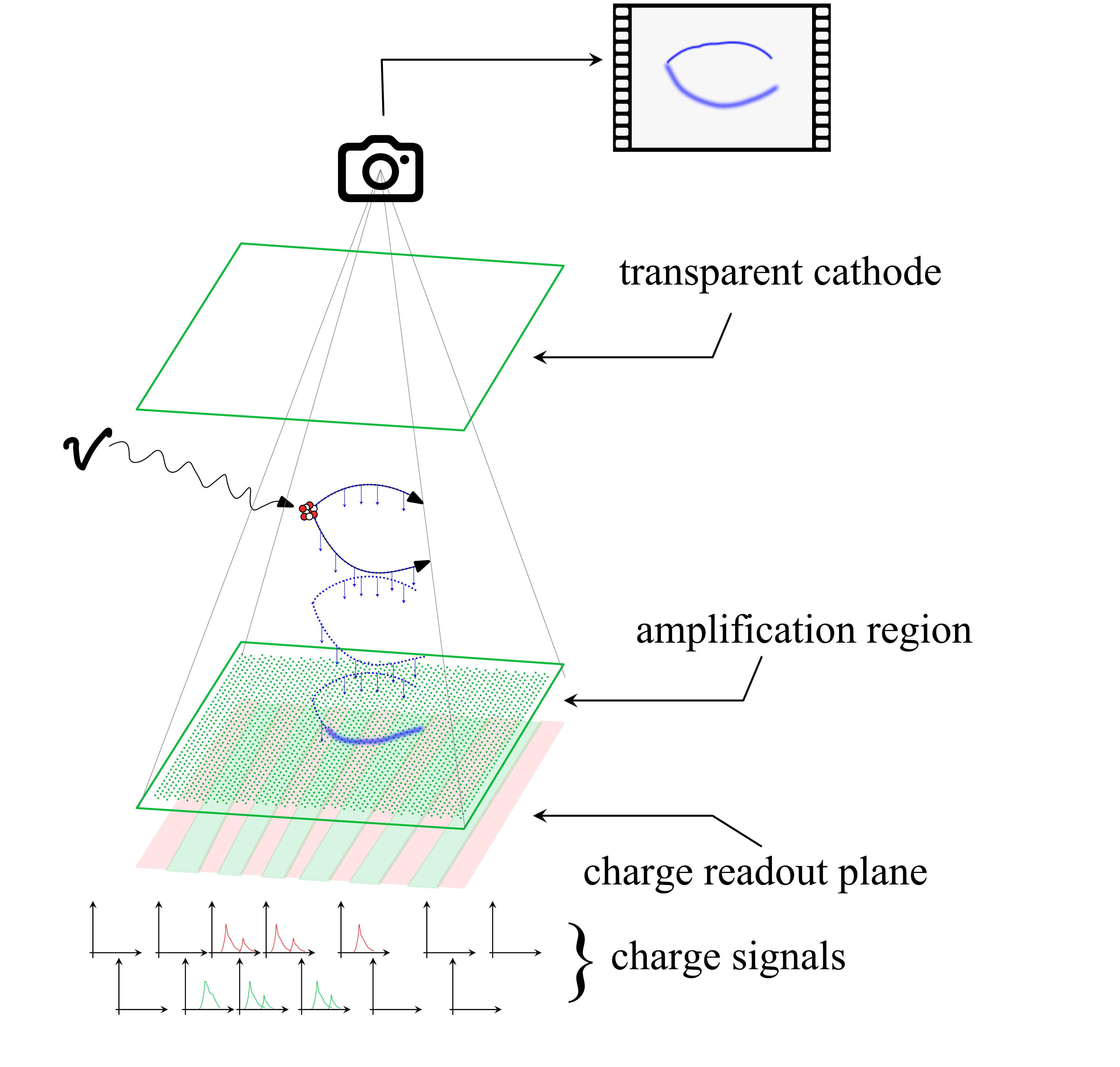}}
\subfloat[]{\label{sec:concept:fig:otpc:gems}\includegraphics[width=0.37\columnwidth, trim= 0 425 0 600, clip=true]{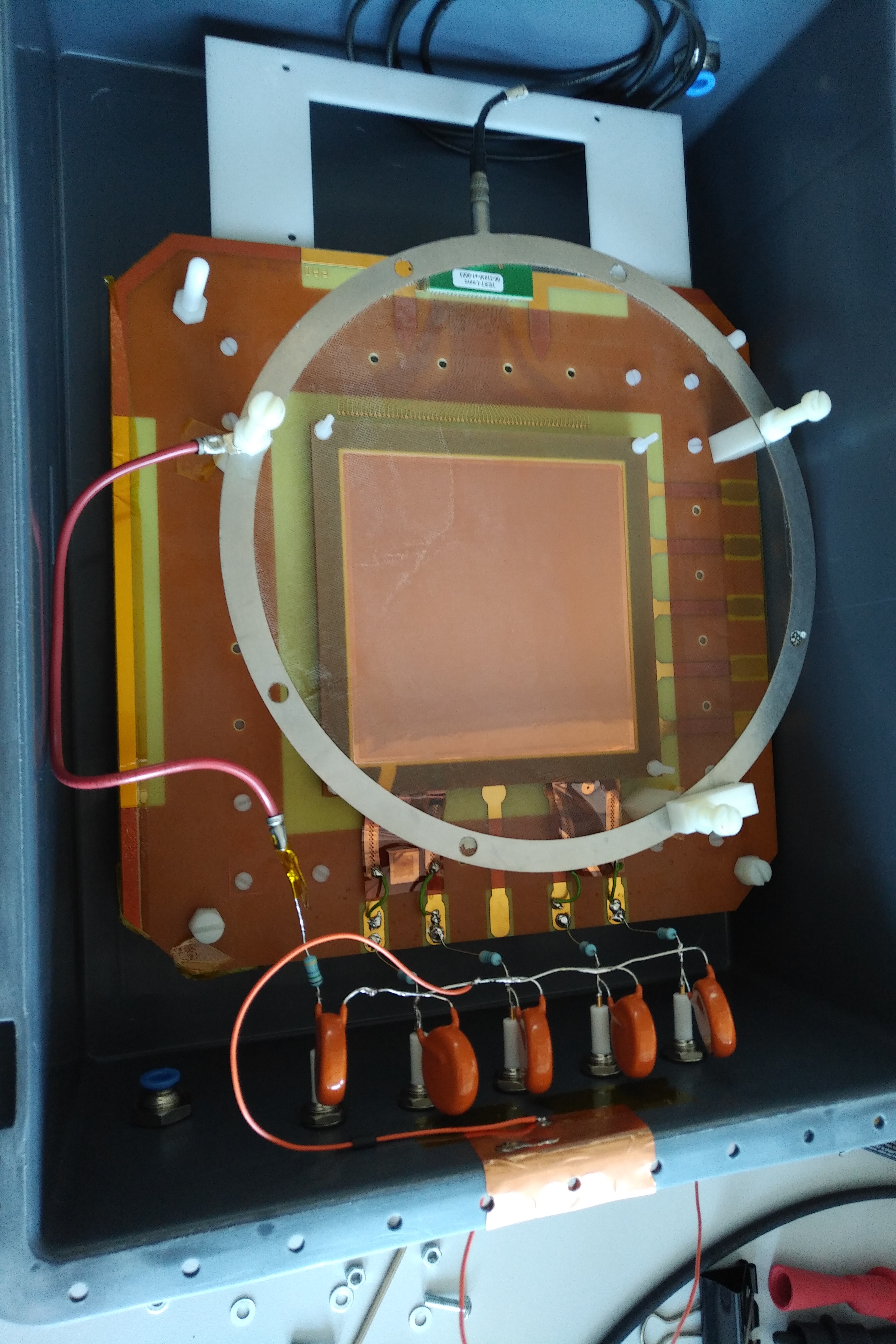}}
\caption{\label{sec:concept:fig:otpc:concept:gems}\protect\subref{sec:concept:fig:otpc:concept} Working principle of a TPC with hybrid optical and charge readout. A neutrino ($\nu$) scatters on a gas atom's nucleus and two charged particles are ejected and create ionisation tracks in the TPC gas, which drift to an amplification stage. A camera images the photons produced during the gas amplification, capturing a 2D projection of the tracks in the drift volume onto the readout plane. Additional charge readout with coarse segments provides the coordinate in the third dimension. \protect\subref{sec:concept:fig:otpc:gems} A photo of the test set-up using a double GEM stack.}
\end{figure}
The idea to read out TPCs by imagining the gas amplification stage with cameras originated in the 1980's, see \cite{CHARPAK1987177,BRESKIN1989457} and the sources therein. \Figref{sec:concept:fig:otpc:concept} explains the principle of a TPC with hybrid optical- and charge readout. A camera is focused onto the gas amplification plane, imaging photons produced during the amplification of the ionisation electrons in the drift volume. The area in the amplification stage imaged by one pixel of a camera can be as small as $\leq100\times\SI{100}{\micro\meter\squared}$, which is more than sufficient to sample several points along a \SI{1}{\centi\meter} track. Cameras are integrated devices where all pixels are read out as one channel, providing a two dimensional (2D) image as the readout object. Conventional charge readout requires many channels to be read out individually and to be reconstructed to create a similar object. The price per channel is also lower for a camera than for a highly segmented charge readout plane.\\
The optical readout measures the projection of tracks onto the readout plane. There are different approaches to measure the coordinate perpendicular to the readout plane, $z$, as \textit{e.g.}: a) The use of a very fast camera to take several images whilst the primary ionisation electrons drift and are amplified \cite{Roberts_2019}. Thus the optical readout itself samples in $z$. b) Readout of the charge signal induced on an anode as a whole \cite{Leyton_2016,BATTAT20146,instruments5020022} or the use of an additional light measurement with \textit{e.g.} a Photo Multiplier Tube (PMT) \cite{Antochi:2020hfw}. The single charge channel or the PMT provides a waveform which allows to draw conclusions about the $z$ position of tracks inside the TPC.\\
We propose a coarsely segmented charge readout plane as sketched in \figref{sec:concept:fig:otpc:concept} providing location information in the readout plane as well as time dependent information. The goal of such a configuration is to provide accurate matching between the optical and the charge readout for events with several tracks. The optimal size of the charge readout segments together with the camera's frame rate will be optimised with a small prototype. We are currently commissioning a set-up with a $10\times\SI{10}{\centi\meter\squared}$ readout plane with 128 rectangular segments and Gas Electron Multipliers (GEMs) \cite{sauli1997gem} as amplification stage. Optical readout with GEMs has already been demonstrated \textit{e.g.} by \cite{Brunbauer:2018nyz}. We plan to group segments of the charge readout together to increasingly larger areas and analyse the performance.

\section{Gas gain simulations}
\label{sec:simulations}

Only a certain fraction on the order of $10^{-4}$ of the original photons reach the camera chip, when focusing a camera with a lens on the amplification region. In order to overcome this factor, enough photons need to be produced, {\it i.e.} the gas amplification factor (gas gain) needs to be large enough. A gas mixture for an optical or hybrid TPC needs thus to be optimised for the gas gain, light emission in an interesting wavelength band, the gas pressure, and diffusion. Furthermore the gas mixture has to suit the physics programme for which the TPC is build. In {\it e.g.} \cite{Hamacher-Baumann:2020ogq} mixtures are examined for their suitability to measure neutrino nucleus scattering.\\
Simulating the gas gain of an amplification stage in \textsc{garfield++} \cite{garfieldpp} can be time intensive, it is thus desirable to derive a functional expression for the gas gain $g(\dots)$ which can be used instead of running new simulations. This function should depend on the gas mixture, the pressure and the applied voltages. As a test case we have simulated the wire plane of an ALICE inner ROC \cite{alme2010alice} for various $\text{Ar}$-$\text{CO}_2$ mixtures. This case may not be particular interesting for optical readout, but it has an application for a high pressure TPC for neutrino physics: There is a plan to use the wire chambers of the ALICE TPC for the DUNE near detector's gas filled TPC \cite{abud2021deep}. Recently, these ROCs were removed from the ALICE TPC to make way for new ones \cite{Adolfsson:2020pcf}.\\
\begin{figure}
\centering
\subfloat[]{\label{sec:concept:fig:otpc:gainana:polya}\includegraphics[width=0.32\columnwidth, trim= 0 0 0 0, clip=true]{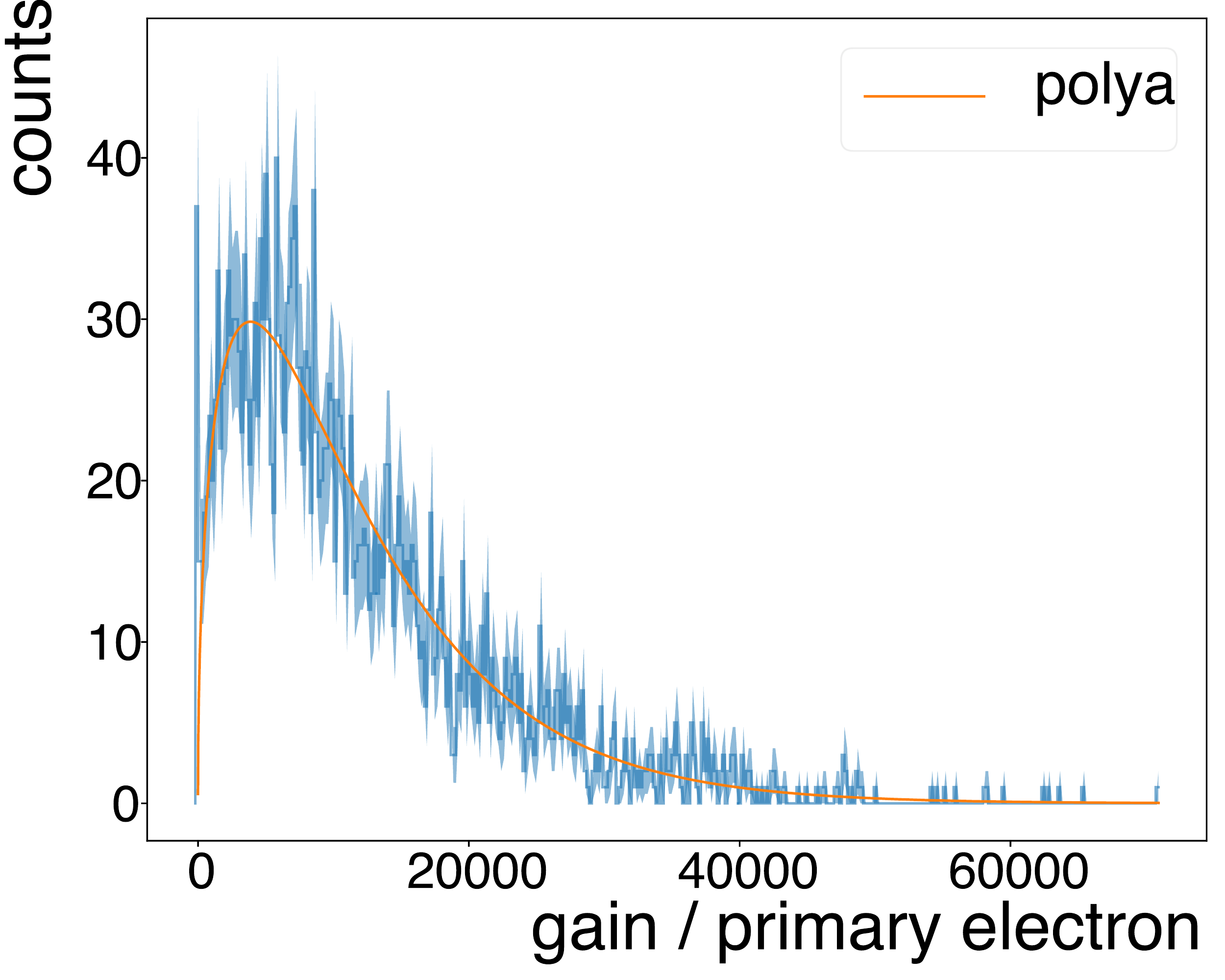}}
\subfloat[]{\label{sec:concept:fig:otpc:gainana:gain}\includegraphics[width=0.32\columnwidth, trim= 0 0 0 0, clip=true]{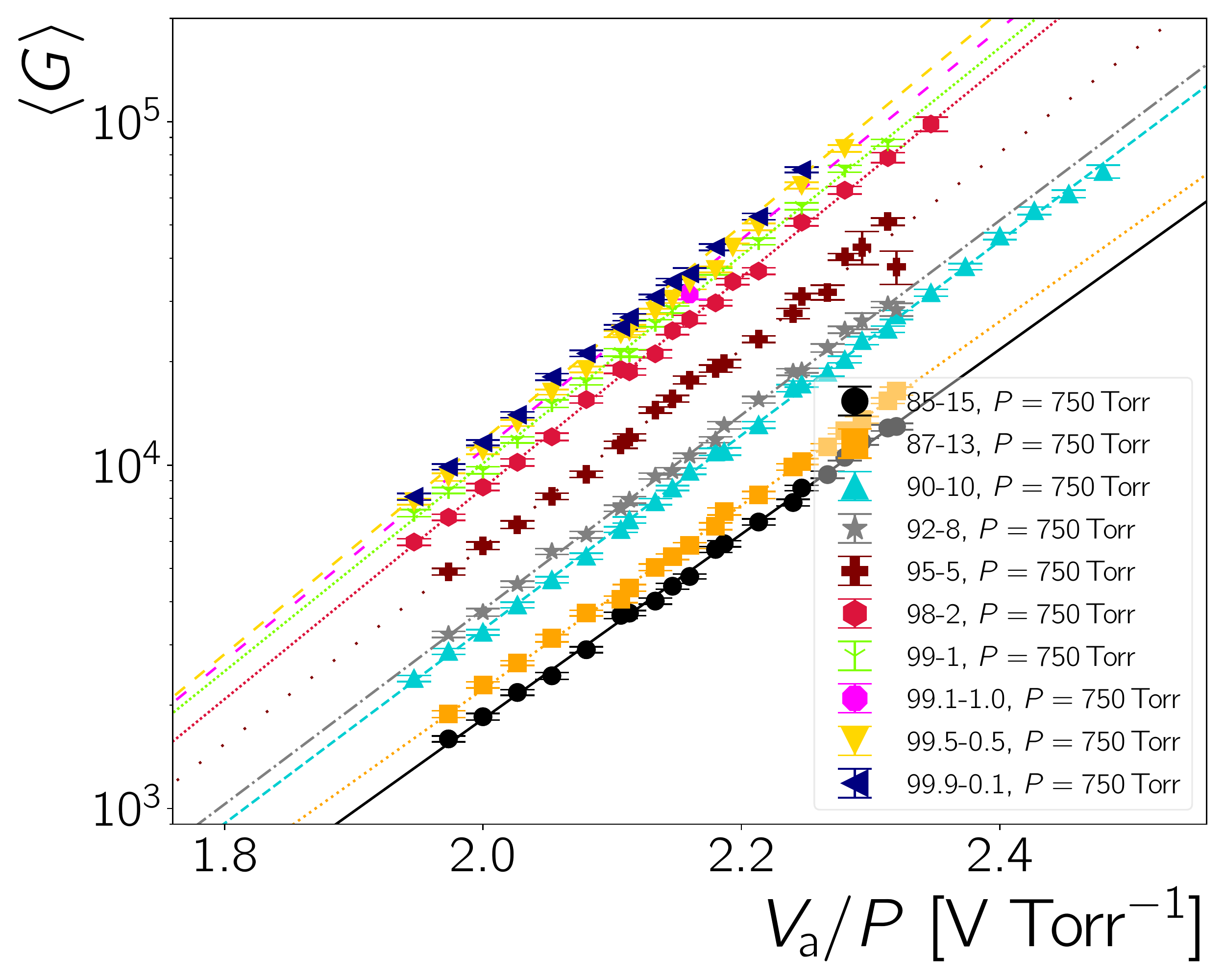}}
\subfloat[]{\label{sec:concept:fig:otpc:gainana:2d:fit}\includegraphics[width=0.32\columnwidth, trim= 0 0 0 0, clip=true]{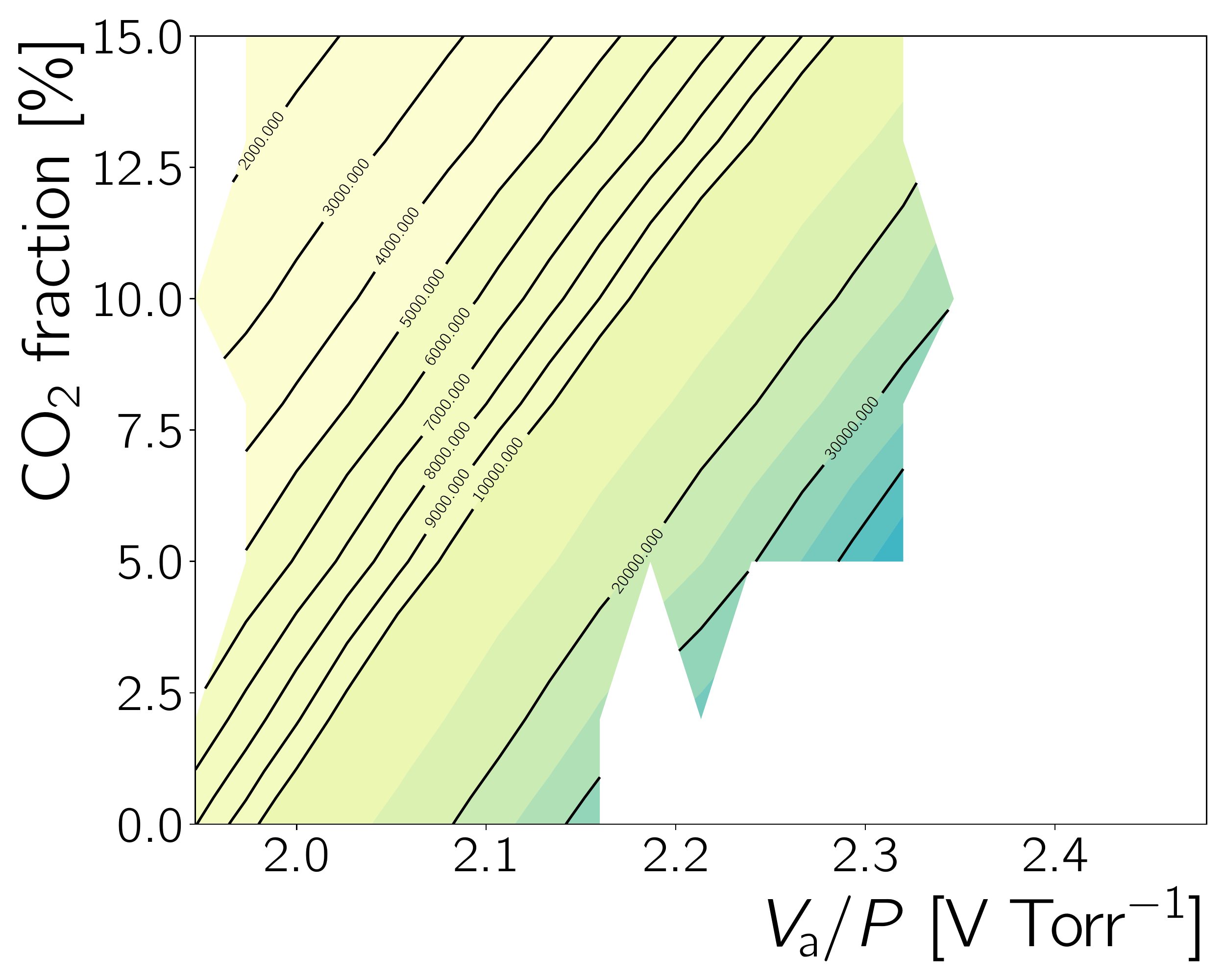}}
\caption{\label{sec:concept:fig:otpc:gainana}\protect\subref{sec:concept:fig:otpc:gainana:polya} Simulation result for a simulation of pure argon at a pressure of \SI{750}{Torr} and an anode voltage of \SI{1500}{\volt}. \protect\subref{sec:concept:fig:otpc:gainana:gain} Gas gain curves for various gas mixtures. \protect\subref{sec:concept:fig:otpc:gainana:2d:fit} A parametrisation of the gas gain (colour in plot) as function of the gas mixture and the reduced anode wire voltage. For more details see the text. Note that the Penning effect is not included in these simulations and the gain values are hence lower than measured ones.}
\end{figure}
In the simulations we place a cluster with 25 electrons at randomised positions above the wire planes. The electrons drift then towards the anode wires and are amplified. The final electron count per primary electron is histogramised for 100 of such simulations (\figrefbra{sec:concept:fig:otpc:gainana:polya}). To extract the gas gain the histogram is fitted by a Polya function:
\begin{align*}
  \mathcal{P}\left( \text{gain} \right) &= \frac{c_{\mathcal{P}}}{\langle G\rangle} \cdot \frac{\left(\theta -1\right)^{\left(\theta-1\right)}}{\Gamma\left(\theta-1\right)} \cdot \left(\frac{\text{gain}}{\langle G\rangle}\right)^{\theta}  \cdot  \exp \left(-\left(\theta-1\right)\cdot \frac{\text{gain}}{\langle G\rangle} \right)
\end{align*}
where $\theta = \left(\langle G\rangle^2-\sigma_{\mathcal{P}}^{2}\right)/\sigma_{\mathcal{P}}^2$, ``gain'' is the gain per primary electron in \figref{sec:concept:fig:otpc:gainana:polya},  and $\langle G\rangle$, $\sigma_{\mathcal{P}}$ and $c_{\mathcal{P}}$ are fit parameters, of which $\langle G\rangle$ is the gas gain. All points in \figref{sec:concept:fig:otpc:gainana:gain} are $\langle G\rangle$ values taken from $\mathcal{P}$ fits to different gas gain histograms. We take $\text{log}\langle G\rangle$ for all gas gain values and fit
\begin{equation}
g \left(\text{CO}_2\ \text{fraction}, \frac{V_{\text{a}}}{P}\right)\; =\; p_0 + p_1 \cdot \text{CO}_2\ \text{fraction} + p_2 \cdot \frac{V_{\text{a}}}{P} + p_3 \cdot \text{CO}_2\ \text{fraction} \cdot \frac{V_{\text{a}}}{P}
\label{sec:simulations:surfacefit}
\end{equation}
as function of the $\text{CO}_2$ fraction and the reduced anode wire voltage $V_{\text{a}}/P$. This choice is based on the $\sim\!\text{exp}\left(\frac{V_{\text{a}}}{P}\right)$ and $\sim\!\text{exp}\left(-\text{CO}_2\ \text{fraction}\right)$ dependence of the gas gain. \Figref{sec:concept:fig:otpc:gainana:2d:fit} shows the fit result with $p_0=-4.208$,  $p_1=6.775$ , $p_2=-\SI{0.0453}{Torr\ \centi\meter\per\volt}$, and $p_3=-\SI{0.0401}{Torr\ \centi\meter\per\volt}$. The $\chi^2/N_{\text{dof}}$ for the fit is $47.2$, indicating that a better function than Equation \eqref{sec:simulations:surfacefit} could be chosen to represent $\text{log}\langle G\rangle$.

\section{Outlook}
Whilst there is certainly still room for improvement, the simulated gain curves show that their dependence on the $\text{CO}_2$ fraction in $\text{Ar}$-$\text{CO}_2$ mixtures and on $V_{\text{a}}$ is smooth enough to find a function to describe the gas gain in the relevant parameter space for \textit{e.g.} a DUNE near detector TPC filled with gas. Full simulations, including the Penning effect, are on the way. In the future these can be compared to measurements done for the DUNE collaboration, too.\\
Furthermore, with easing of access restrictions to the laboratory, work on a prototype amplification stage for a hybrid readout TPC is on the way.

\section{Acknowledgements}

This project has received funding from the European Union’s Horizon 2020 Research and Innovation programme under grant agreement No 101004761.

% Bibliography
\section*{References}
\bibliographystyle{iopart-num}
\bibliography{proceedings_bib}

\end{document}